\documentclass[prb,twocolumn,amsmath,amssymb,showpacs]{revtex4}
\usepackage{graphicx}

\def\beq{\begin{eqnarray}}
\def\eeq{\end{eqnarray}}
\def\beqa{\begin{eqnarray}}
\def\eeqa{\end{eqnarray}}

\begin{document}

\title{
One-particle spectral properties of the $t$-$J$-$V$ model on the triangular 
lattice near charge order
}

\author{M. Bejas\dag, A. Greco\dag, A. Muramatsu\ddag \; and, A. Foussats\dag}
\affiliation{
{\dag} Facultad de Ciencias Exactas, Ingenier\'{\i}a y Agrimensura and Instituto de F\'{\i}sica Rosario (UNR-CONICET).
Av. Pellegrini 250-2000 Rosario-Argentina.\\
{\ddag} Institut f\"ur Theoretische Physik III, Universit\"at Stuttgart, Pfaffenwaldring 57, D-70550 Stuttgart, Germany.
}

\date{\today}

\begin{abstract}
We study the $t$-$J$-$V$ model beyond mean field level at finite doping
on the triangular lattice. The Coulomb repulsion $V$ between nearest 
neighbors brings the system to a charge 
ordered
state for $V$ larger than a critical value $V_c$.
One-particle spectral properties as self-energy, spectral functions and the 
quasiparticle weight are studied near and far from the charge 
ordered
phase.
When the system approaches the charge 
ordered
state, charge fluctuations become 
soft and they strongly influence the system
leading to 
incoherent 
one-particle excitations.
Possible implications for cobaltates are given.
\end{abstract}

\pacs{71.10.Fd, 71.27.+a, 71.45.Lr}

\maketitle

\section{Introduction}
Strongly correlated electronic systems continue to be at the center of
interest
in condensed matter, the $t$-$J$ model being a paradigm for 
theoretical studies.
Since the early days of high-$T_c$ superconductivity the $t$-$J$ model on the 
triangular lattice attracted a great deal of attention, since it was believed
to be
a candidate for realizing the resonating valence bond scenario 
\cite{anderson87}.

Recently, the interest about electronic correlations on a two dimensional 
triangular lattice received a new motivation.
Since superconductivity was discovered in hydrated cobaltates 
\cite{takada03} ($Na_xCoO_2.yH_2O$)  an enormous amount of attention has been 
focused on this system.
Cobaltates are $3d$-electron systems having a quasi-two-dimensional structure 
of $CoO_2$ layers, where the $Co$-ions are located on a triangular lattice.
The interplay 
between electronic correlations and the proximity 
to a charge density wave
were
proposed 
recently as 
important ingredients
for describing superconductivity in 
such materials \cite{motrunich04,tanaka04,foussats05,foussats06}. 
This proposal was motivated by various
experimental reports which suggest the proximity of the system to charge 
ordering \cite{shimojima05,lemmens05,qian06}.
A minimal model for strongly correlated electronic systems close to charge 
ordering is the $t$-$J$-$V$ model on the triangular lattice,
where $V$ is the Coulomb interaction between nearest neighbor sites.
The parameter $V$ is responsible for triggering the charge order instability.

In this paper we concentrate on the consequences of the proximity to 
charge ordering on one-particle spectral properties. For this study we use 
a recently developed large-$N$ approach \cite{foussats02,bejas06} that leads
to a systematic treatment of fluctuations beyond the mean-field level, 
allowing thus the calculation of the self-energy. 
Related expansions were used in the past in the frame of a slave boson 
formulation for the t-J model \cite{wang92}. However, due to the gauge-fields 
inherent to slave bosons, such an expansion cannot be performed in a 
systematic way \cite{lee06}. 
Our formalism is based on  $X$-operators that do not introduce gauge-fields, 
such that the present treatment is free of the above mentioned
difficulties. 
In general, large $N$ expansions give preference to one kind of fluctuations 
over others. In our case, the $1/N$ expansion (with the same $N=\infty$ limit 
as in slave bosons) gives preference to charge fluctuations over magnetic 
ones. Therefore, it fails to describe the limit of zero doping of an 
antiferromagnet on a square lattice, where antiferromagentic long-range 
order should be present, a difficulty that is shared also by the U(1) slave 
boson formulation \cite{lee06}.
However, in the present case we focus on rather high doping ($\sim$ 30\%)
on a frustrated lattice, such that in fact, charge fluctuations are the
dominant factor. For the square lattice, as discussed in 
Ref.[\onlinecite{bejas06}] our method is better for large doping, 
corresponding to the overdoped region of high-$T_c$ cuprates.

Our results show how, starting from an already reduced value by 
correlation effects, the quasiparticle weight vanishes on approaching the 
charge ordering instability, leading to a redistribution of the one-particle 
spectral weight. Furthermore, the soft charge modes responsible for the 
phenomena above are identified.

The paper is organized as follows. In Sec.\ II, a summary of the formalism and
details of the self-energy calculation are given.
In Sec.\ III, the charge instability is studied. In Sec.\ IV,
results on one-particle spectral properties (self-energy,
quasiparticle weight, spectral functions) are presented.
In Sec.\ V, possible implications for cobaltates are discussed. Section VI 
presents
conclusion and discussions.

\section{ The model and the theoretical framework}

The $t$-$J$-$V$ model on the triangular lattice
is
given by
\begin{eqnarray}\label{eq:Hc}
H &=& -t \sum_{<ij>,\sigma} (\tilde{c}^\dag_{i\sigma}
\tilde{c}_{j\sigma}+h.c.)\nonumber \\
&+& J \sum_{<ij>} (\vec{S}_i \vec{S}_j-\frac{1}{4} n_i n_j)
+ V \sum_{<ij>} n_i n_j \; ,
\end{eqnarray}
where $t$, $J$ and $V$ are the  hopping, the exchange interaction
and the Coulomb repulsion, respectively, between nearest-neighbor sites 
denoted by $\langle ij \rangle$.
$\tilde{c}^\dag_{i\sigma}$ and $\tilde{c}_{i\sigma}$ are the
fermionic creation and destruction operators, respectively, under the 
constraint that double occupancy is
excluded, and $n_i$ is the corresponding density operator at site $i$.

As described in Ref.\ [\onlinecite{foussats06}] the Hamiltonian (\ref{eq:Hc}) 
can be written in terms of Hubbard operators\cite{hubbard63} as:
\begin{eqnarray}
H &=& - \frac{t}{N}\sum_{<ij>,p}\;(\hat{X}_{i}^{p 0}
\hat{X}_{j}^{0
p} + h.c.) \nonumber \\
&+& \frac{J}{2N} \sum_{<ij>;pp'} (\hat{X}_{i}^{p p'}
\hat{X}_{j}^{p' p} - \hat{X}_{i}^{p p} \hat{X}_{j}^{p' p'})
 \nonumber\\
&+& \frac{V}{N}\sum_{<ij>;p p'} \hat{X}_{i}^{p p} \hat{X}_{j}^{p' p'}
-\mu\sum_{i,p}\;\hat{X}_{i}^{p p} \; , \label{eq:H}
\end{eqnarray}
where, in addition, in order to perform a large-$N$ expansion, the 
spin index $\sigma$ was
extended to a new index $p$ running from $1$ to $N$.
In order to obtain a finite theory in the $N$-infinite limit, we rescaled 
$t$, $J$ and $V$ as $t/N$, $J/N$ and $V/N$, respectively.
We included 
in eq.\ (\ref{eq:H}) 
the chemical potential $\mu$.
The operators $\hat{X}^{pp'}$ are boson-like while $\hat{X}^{0p}$ and 
$\hat{X}^{p0}$ are fermion-like \cite{hubbard63}.

\begin{figure}
\vspace{1cm}
\begin{center}
\setlength{\unitlength}{1cm}
\includegraphics[width=8cm,angle=0]{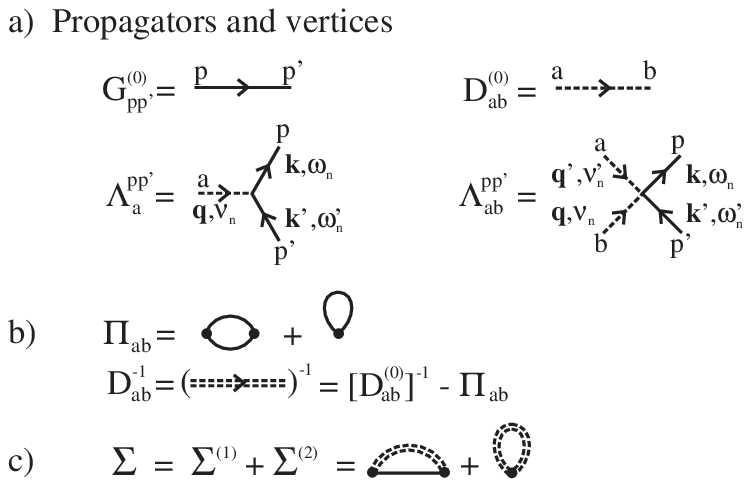}
\end{center}
\caption{Summary of the Feynman rules. a) Solid lines represent the
propagator $G^{(0)}$ (eq.\ (\ref{G0})). Dashed lines represent the 
$8 \times 8$ boson propagator
$D^{(0)}$ (eq.\ (\ref{D0})) for the $8$-component field $\delta X^a$. 
$\Lambda^{pp'}_a$ and $\Lambda^{pp'}_{ab}$ represent
the interaction between two fermions $f_p$ and one or two bosons $\delta X^a$ 
respectively. b) Irreducible boson self-energy $\Pi_{ab}$. Double dashed 
lines correspond to a dressed boson propagators.
c) Contributions to the electron self-energy $\Sigma({\bf k},\omega)$ 
in
${\cal O}(1/N)$.
}
\label{fig:FR}
\end{figure}

In Ref.\ [\onlinecite{foussats06}] the large-$N$ formalism was described in 
detail and here a short
summary is presented which will be useful for the 
calculation of the self-energy and the spectral function.
The Feynman rules of the method are summarized in Fig.\ \ref{fig:FR}.

To leading order in  $1/N$, we associate with the $N$-component fermion field 
$f_{p}$ a propagator connecting two generic components $p$ and $p'$ (solid 
line in Fig.\ref{fig:FR})
\begin{eqnarray}\label{G0}
G^{(0)}_{pp'}({\bf k}, \omega_{n}) = - \frac{\delta_{pp'}}{i\omega_{n} - E_{k}}
\end{eqnarray}
which is of ${\cal O}(1)$ and where $E_{k}$ is
\begin{eqnarray}\label{Ek}
E_{k} = -2(t r_0+\Delta) (\cos
k_{x}+2\cos\frac{k_{x}}{2}\cos\frac{\sqrt{3}}{2}k_{y} ) - \mu \; .
\end{eqnarray}
${\bf k}$ and $\omega_{n}$ are the momentum and the fermionic  Matsubara 
frequency of the fermionic field, respectively.
The fermion variables $f_{ip}$ are proportional to the $\hat{X}$-operators 
($f_{ip}=\frac{1}{\sqrt{Nr_0}} \hat{X}^{0p}_i$ ) and 
are not 
associated 
with the spinons from the slave boson approach.

The mean-field values $r_0$ and $\Delta$ must be determined by minimizing the 
leading order theory. From the completeness
condition\cite{foussats06}, $\hat{X}^{00}+\sum_p \hat{X}^{pp} = \frac{N}{2}$, 
$r_0$ is equal to $x/2$ where $x$ is the
doping away from half-filling. On the other hand, the expression for $\Delta$ 
is
\beq{\label {Delta}}
\Delta = \frac{J}{2N_s} \frac{1}{3}\sum_{k \eta} \cos(k_\eta) n_F(E_k) \; ,
\eeq
where $n_F$ is the Fermi function and $N_s$ is the number of 
lattice sites. 
For a given doping $x$; $\mu$ and $\Delta$ must be determined 
self-consistently from $(1-x)=\frac{2}{N_s} \sum_{k} n_F(E_k)$ and 
eq.\ (\ref{Delta}).
$k_{\eta}$ is the proyection of ${\bf k}$ over the different bond directions 
${\eta}_{1}=(1,0)$, ${\eta}_{2}=(\frac{1}{2},\frac{\sqrt{3}}{2})$ and 
${\eta}_{3}=(-\frac{1}{2},\frac{\sqrt{3}}{2})$ of the triangular lattice.

We associate with the eight component boson field $\delta X^{a} = (\delta
R\;,\;\delta{\lambda},\; r^{{\eta}_{1}},\;r^{{\eta}_{2}}
,\;r^{{\eta}_{3}},\; A^{{\eta}_{1}},\;
A^{{\eta}_{2}},\;A^{{\eta}_{3}}) $, the inverse of the propagator, connecting 
two generic components a and b (dashed line in Fig.\ \ref{fig:FR}),
\begin{widetext}
\begin{eqnarray}
D^{-1}_{(0) ab}({\mathbf{q}},\nu_{n})= N \left(
 \begin{array}{cccccccc}
 \gamma_{q} &x/2 & 0&0&0&0 &0& 0 \\
   x/2 & 0 & 0 &0& 0 & 0 &0& 0 \\
   0 & 0 & \frac{4}{J}\Delta^{2} & 0&0 & 0&0 & 0 \\
   0 & 0 & 0 & \frac{4}{J}\Delta^{2} & 0 &0&0& 0 \\
   0 & 0 & 0 & 0& \frac{4}{J}\Delta^{2} &0&0& 0 \\
   0 & 0 & 0 & 0 & 0 & \frac{4}{J}\Delta^{2}&0&0 \\
   0 & 0 & 0 & 0 & 0 &0& \frac{4}{J}\Delta^{2}&0 \\
   0 & 0 & 0 & 0& 0 & 0&0&\frac{4}{J}\Delta^{2} \
 \end{array}
\right), \label{D0}
\end{eqnarray}
\end{widetext}
where $\gamma_{q}=(2V-J)(x/2)^{2}\; \sum_\eta \cos k_{\eta}$ 
\cite{aclara1} and the indices $a$, $b$
run from 1 to 8.
${\bf q}$ and $\nu_{n}$ are the momentum and the Bose Matsubara frequency of 
the boson field, respectively.

The first component $\delta R$ of the $\delta X^a$ field is connected with 
charge fluctuations via $\hat{X}^{00}_i=Nr_0(1+\delta R_i)$, where 
$\hat{X}^{00}_i$ is the Hubbard operator associated with the number of 
vacancies at site $i$.
$\delta \lambda$ is the fluctuation of the 
Lagrange
multiplier $\lambda_i$ 
associated with the completness condition. $r_i^{\eta}$ and $A_i^{\eta}$ 
correspond, respectively, to the amplitude and the phase fluctuation of the 
bond variable $\Delta_i^{\eta} = \Delta \, (1 + r_i^{\eta} + i \, A_i^{\eta})$.

Fermions $f_p$ interact with the boson $\delta X^a$ via three and four leg 
vertices, namely $\Lambda^{pp'}_{a}$ and $\Lambda^{pp'}_{ab}$ respectively.
The explicit expressions for these vertices can be found in 
Ref.\ [\onlinecite{foussats06}].

Each vertex conserves momentum and energy and they are of ${\cal O}(1)$.
In each diagram there is a minus sign for each fermion loop and
a topological factor.

The bare boson propagator $D_{(0)ab}$ (the inverse of
eq.\ (\ref{D0})) is ${\cal O}(1/N)$. From the Dyson equation, $D_{ab}^{-1}
= D_{(0) ab}^{-1} - \Pi_{ab}$, the dressed components $D_{ab}$
(double dashed line in Fig.\ \ref{fig:FR}b) of the boson propagator can
be found after the evaluation of the $8 \times 8$ boson
self-energy matrix $\Pi_{ab}$. Using the Feynman rules,
$\Pi_{ab}$ can be evaluated through the diagrams of Fig.\ref{fig:FR}b as
 shown in Ref.\ [\onlinecite{foussats06}].

This formalism is used 
here
for calculating
self-energies and one-particle spectral functions\cite{bejas06}.
The Green's function (\ref{G0}) corresponds to the $N$-infinite
propagator which includes no dynamical corrections; these appear
at higher order in the $1/N$ expansion. Using the
Feynman rules, the total self-energy in ${\cal O}(1/N)$ is obtained
adding the contribution of the two diagrams shown in 
Fig.\ \ref{fig:FR}c.

After performing the Matsubara sum and the
analytical continuation $i\omega_n=\omega+i\xi$,
the imaginary part of $\Sigma$ is
\begin{eqnarray}\label{eq:SigmaIm}
    Im \, \Sigma({\mathbf{k}},\omega) &=& \frac{1}{2 N_{s}}
\sum_{{\mathbf{q}}} h_{a}(k,q,\omega-E_{{k-q}}) \nonumber\\
&& B^{ab}({\mathbf{q}},\omega-E_{{k-q}}) \; h_{b}(k,q,\omega-E_{{k-q}}) 
\nonumber\\
&&\times [n_{F}( -E_{k-q}) +n_{B}(\omega-E_{{k-q}})]
\end{eqnarray}
\noindent where $n_B$ is the Bose factor, $B_{ab}$ is
\begin{equation}\label{B}
B^{ab}({\mathbf{q}},\nu)=-2\lim_{\eta\rightarrow0}\; Im \, [D_{ab}
({\mathbf{q}},i\nu_{n}\rightarrow\nu+i\eta)].
\end{equation}
\noindent and the 8-component vector \cite{aclara2} $h_{a} (k,q,\nu)$ is
\begin{widetext}
\begin{eqnarray}\label{eq:ha}
h_{a} (k,q,\nu) &=& \left(
\frac{2E_{k-q}+\nu+2\mu}{2} 
+ 2\Delta \sum_{\eta} \cos(k_\eta-\frac{q_\eta}{2}) \cos(\frac{q_\eta}{2})
;1
; -2 \Delta\; \cos(k_{{\eta}_{1}}-\frac{q_{{\eta}_{1}}}{2})
; -2 \Delta\; \cos(k_{{\eta}_{2}}-\frac{q_{{\eta}_{2}}}{2}); \right. 
\nonumber \\
&& \left.
-2 \Delta\; \cos(k_{{\eta}_{3}}-\frac{q_{{\eta}_{3}}}{2})
; 2 \Delta\; \sin(k_{{\eta}_{1}}-\frac{q_{{\eta}_{1}}}{2})
; 2 \Delta\; \sin(k_{{\eta}_{2}}-\frac{q_{{\eta}_{2}}}{2})
; 2 \Delta\; \sin(k_{{\eta}_{3}}-\frac{q_{{\eta}_{3}}}{2}) \right) \;.
\end{eqnarray}
\end{widetext}

It is interesting to show $Im \, \Sigma$ expliciting the terms $B_{RR}$, 
$B_{\lambda R}$ and $B_{\lambda \lambda}$:
\begin{eqnarray}\label{eq:SigmaIm0}
Im \, \Sigma({\mathbf{k}},\omega)&=& \frac{1}{2 N_{s}}
\sum_{{\mathbf{q}}} \left\{ \Omega^{2} \;
 B^{RR}({\mathbf{q}},\omega-E_{{k-q}}) \right. \nonumber\\
&& \hspace{-1cm} + \; 2\;\Omega \;
B^{\lambda R}({\mathbf{q}},\omega-E_{{k-q}})
+ \left.
B^{\lambda \lambda}({\mathbf{q}},\omega-E_{{k-q}}) \right\} \nonumber\\
&& \hspace{-1cm}\times \left[n_{F}(-E_{k-q}) + n_{B}(\omega-E_{{k-q}})\right]
+ f({\bf k}, \omega)\; , \nonumber \\
\end{eqnarray}
\noindent where $\Omega=(E_{k-q}+\omega+2\mu)/2 
+ 2\Delta \sum_{\eta} \cos(k_\eta-q_\eta/2) \cos(q_\eta/2)$.
In the function $f({\bf k}, \omega)$ the other terms of eq.\ 
(\ref{eq:SigmaIm}) have been included.
After performing numerical calculations for the parameters used in the 
present paper, $f({\bf k}, \omega)$ results at least an order of magnitud 
lower than the other terms of eq.\ (\ref{eq:SigmaIm0}).

Using the Kramers-Kronig relations, $Re \, \Sigma({{\bf{k}}},\omega)$ can be 
determined from $Im \, \Sigma({{\bf{k}}},\omega)$,
eq.\ (\ref{eq:SigmaIm}) 
and the 
spectral function
$A({\bf k},\omega)=-\frac{1}{\pi} Im \, G({\bf k},\omega)$
can be computed as
\begin{eqnarray} \label{A}
A({\bf k},\omega)= -\frac{1}{\pi}
\frac{ Im \, \Sigma({\bf k},\omega) }
{ [\omega - E_{\bf k} - Re \, \Sigma({\bf k},\omega)]^2  
+  [Im \, \Sigma({\bf k},\omega)]^2 }\nonumber \\
\end{eqnarray}

Notice that the self-energy is calculated using the propagator $G({\bf
k},\omega)$ for the X-operators, that are proportional to the fermionic
$f$-operators, which cannot be related to usual fermions.

\section{Charge density wave instabilities} \label{sec:CDW}
The inclusion of a nearest neighbors Coulomb repulsion $V$ favors a charge 
density wave (CDW) state.
In this section we will discuss the charge instability\cite{foussats06} which 
will be of interest for analysing self-energy corrections.

The charge-charge correlation function $\chi^c_{ij}$ can be written as 
\cite{foussats02,gehlhoff95}
\begin{eqnarray}\label{chi}
\chi^c_{ij}(\tau)=\frac{1}{N} \sum_{pq} <T_\tau X_i^{pp}(\tau)
X_j^{qq}(0)> \; .
\end{eqnarray}

Using the completeness condition and the relation between $\hat{X}^{00}_i$ 
and $\delta R_i$ ($\hat{X}^{00}=Nr_0(1+\delta R_i)$),
in Fourier space,
\begin{eqnarray}\label{chiv1}
\chi^{c}({\bf{q}},\omega)=-N \left(\frac{x}{2}\right)^2
D_{RR}({\bf{q}},\omega) \; .
\end{eqnarray}
Then, in our formulation, the charge correlation function is proportional to
the component $(1,1)$ (also called $D_{RR}$) of the dressed boson propagator 
$D_{ab}$.

The divergence of the static charge susceptibility (\ref{chiv1}) marks the 
onset for the CDW instability.
For the triangular lattice the charge susceptibility diverges at 
$
{\bf Q}=(4/3\pi,0)$ for 
$V=V_c$.
This new phase is called $\sqrt{3} \times \sqrt{3}$ CDW 
\cite{motrunich04,foussats05,foussats06}.
In what follows all energies ($V$, $J$, etc.) are given in units of $t$.

\begin{figure}
\begin{center}
\setlength{\unitlength}{1cm}
\includegraphics[width=8.5cm,angle=0.]{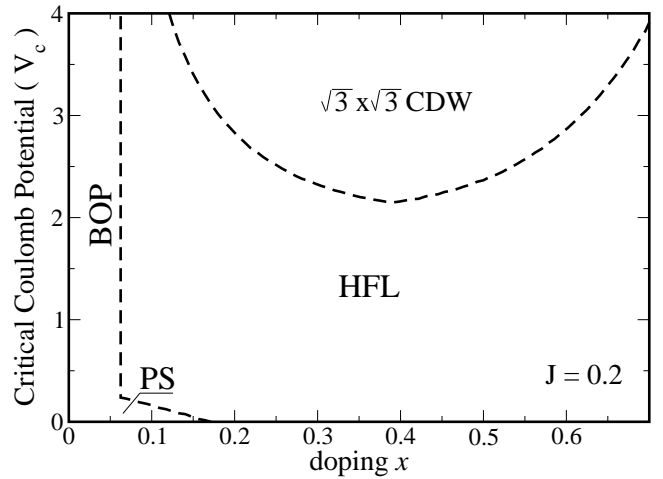}
\end{center}
\caption{
Phase diagram in the $V_c-x$ plane for $J=0.2$. The homogeneous Fermi liquid 
(HFL) is unstable against
phase separation (PS), bond order phase (BOP) and the 
$\sqrt{3} \times \sqrt{3}$ CDW. BOP and PS are discussed
in Ref.\ [\onlinecite{foussats06}] and are presented here 
only for a complete 
description of the phase diagram. For a given
doping $x$ the static charge susceptibility diverges at $V_c$ where the onset 
to a
$\sqrt{3} \times \sqrt{3}$ CDW occurs.
}
\label{fig:fase}
\end{figure}

Figure \ref{fig:fase} shows the phase diagram in the $V_c-x$ plane 
at temperature $T=0$.
As in Ref.\ [\onlinecite{foussats06}], we will use $J=0.2$. 
The different phases are
the homogeneous Fermi liquid (HFL),
the $\sqrt{3} \times \sqrt{3}$ CDW, 
the bond-order phase (BOP), and a phase separation (PS) region. 
However, in this paper
we focus
only on 
the proximity to the HFL-CDW transition
(see Ref.\ [\onlinecite{foussats06}] for details about the BOP and PS phases).
For the large doping 
studied here 
results are very robust against different $J$ values 
as long as 
they are not considered to be unphysically large. 
Only for unphysical large values $J \sim 1$,
the
BOP and PS regions approach 
the doping levels considered here. But for $J < 0.5$, 
a value well beyond the physical ones in e.g.\ cobaltates,
the HFL and CDW phases are well defined beyond $x \sim 0.15$.

\begin{figure}
\begin{center}
\setlength{\unitlength}{1cm}
\includegraphics[width=8.5cm,angle=0.]{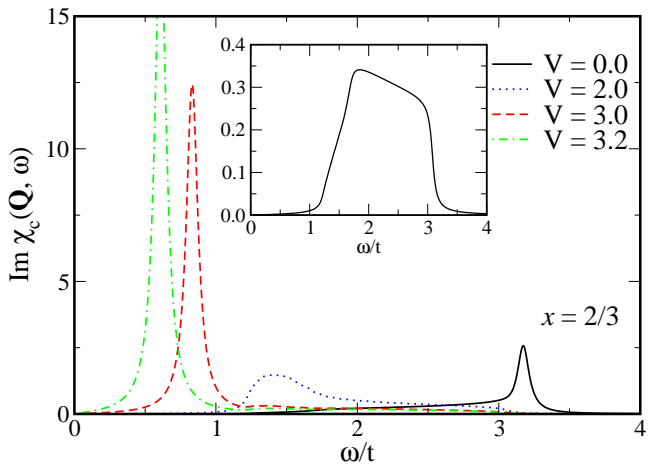}
\end{center}
\caption{(color online)
$Im \, \chi_c({\bf Q},\omega)$ for $x=2/3$, $J = 0.2$ and for different values of $V$ approaching $V_c
\sim 3.6
$.
${\bf Q}=(4\pi/3,0)$, where the static charge susceptibility diverges.
For $V=0$, the collective charge peak is formed at the top
of the particle-hole continuum (inset). With increasing $V$ the collective peak becomes soft
and emerges from the bottom of the continuum accumulating spectral weight.
For $V = V_c \sim 3.6$ this soft collective peak reaches $\omega = 0$ freezing the $\sqrt{3}
\times \sqrt{3}$ CDW phase.
}
\label{fig:sus66}
\end{figure}

\begin{figure}
\begin{center}
\setlength{\unitlength}{1cm}
\includegraphics[width=8.5cm,angle=0.]{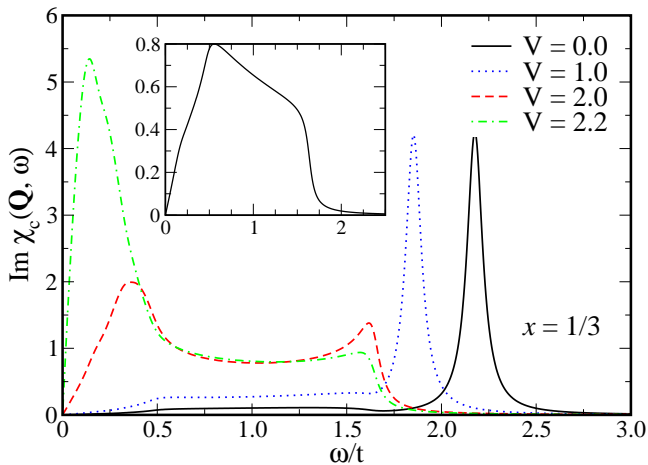}
\end{center}
\caption{(color online)
Similar to Fig.\ \ref{fig:sus66} for $x=1/3$ 
($V_c \sim 2.35$).
Notice the different behaviour with respect to $x=2/3$ (see text).
Inset: particle-hole continuum.
}
\label{fig:sus33}
\end{figure}

Next, we study 
the
charge dynamics 
on
approaching the $\sqrt{3} \times \sqrt{3}$ CDW 
from the HFL.
In Figs.\ \ref{fig:sus66} and \ref{fig:sus33} results for 
$Im\,\chi^c({\bf Q},\omega)$ are presented
for two commensurate doping values $x=2/3$ and $x=1/3$, respectively.
The momentum was fixed at ${\bf Q}=(4/3\pi,0)$.

For $x=2/3$ (Fig.\ \ref{fig:sus66}) and for $V=0$ (solid line), we clearly 
see a
collective peak at $\omega\sim 3.1t$ at the top of the particle-hole 
continuum (inset).
With increasing $V$ the collective peak becomes soft and accumulates spectral 
weight as shown for $V=2.0$ (dotted line),
$V=3.0$ (dashed line) and $V=3.2$ (dotted dashed line).
At $V=V_c \sim 3.6$, the collective peak reaches zero frequency 
condensing
a 
$\sqrt{3} \times \sqrt{3}$ CDW phase.
Therefore, 
the
charge instability can clearly be seen as the softening of the 
collective charge mode.
For $x=2/3$, charge dynamics is very similar to the case of the square 
lattice at one quarter filling 
where the softening of the charge mode was obtained and discussed in the 
context of organic materials\cite{merino03}.

For $x=1/3$ (Fig.\ \ref{fig:sus33}), the situation is somewhat different.
For $V=0$ (solid line) the collective peak is located at $\omega \sim 2.2t$,
above the particle-hole continuum.
With increasing $V$ 
towards
$V_c \sim 2.35$, 
a transfer of
spectral weight 
takes place over a large energy range
since
the collective charge 
mode 
spreads over the particle-hole continuum.

The reason for the different behavior between $x=1/3$ and $x=2/3$ 
is due to
the form of the particle-hole continuum for each doping (insets in 
Figs.\ \ref{fig:sus66} and \ref{fig:sus33}).
For $x=2/3$ there is a gap at low energy, since for $x=2/3$ 
$2{\bf k_F} < {\bf Q}$, 
and
particle-hole transitions are not possible for  
$\omega<1$.
For $V=0$ the collective peak is formed at the top of this continuum 
(Fig.\ \ref{fig:sus66} solid line).
When V increases, the collective peak softens (Fig.\ \ref{fig:sus66}) and for 
large $V$, near $V_c$, emerges from the continuum
until it reaches $\omega=0$ exactly at $V_c$ (see for instance solid 
and dashed lines in Fig.\ \ref{fig:sus66}).
In contrast, for $x=1/3$ there is no gap in the  particle-hole continuum
(inset Fig.\ \ref{fig:sus33}).
When the collective mode enters the particle-hole continuum,
a spread of spectral weight over a wide range takes place, and for $V$
approaching $V_c$, it emerges from the continuum as a resonance.

\begin{figure}
\begin{center}
\setlength{\unitlength}{1cm}
\includegraphics[width=8.5cm,angle=0.]{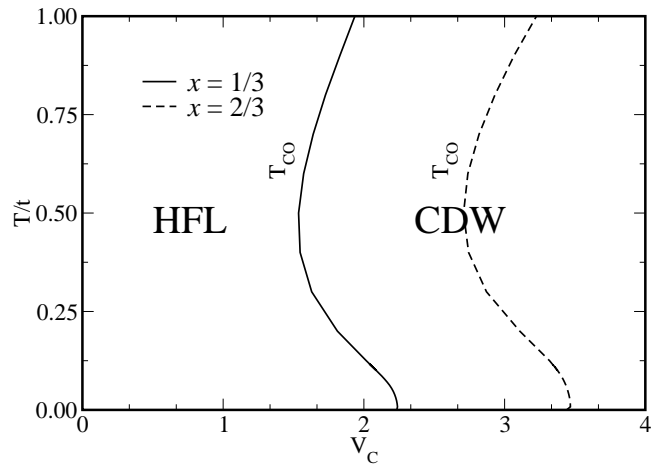}
\end{center}
\caption{
$T$-$V$ phase diagram for $x=1/3$ (solid line) 
and $x=2/3$ (dashed line).
}
\label{fig:TV}
\end{figure}

Before closing the section, we present
in Fig.\ \ref{fig:TV} the $T$-$V$ phase diagram for $x=1/3$ and $x=2/3$.
$T_{CO}$ is the temperature for charge ordering. 
The obtained reentrant behavior of the charge order transition was also 
predicted by 
several calculations \cite{pietig99,hellberg01,hoang02,merino06,merino01}.
 Recently, it was proposed that a reentrant behavior near charge order
is necessary for describing anomalous optical features in $1/4$-filling 
organic materials\cite{merino06} (see Sec. V for more details).

In the next section, we will study self-energy corrections approaching the 
$\sqrt{3} \times \sqrt{3}$ CDW phase and discuss the role played by the soft 
modes mentioned above.

\section{One-particle properties beyond mean-field level} \label{sec:OPEP}
\subsection{Quasiparticle weight}
\begin{figure}
\begin{center}
\setlength{\unitlength}{1cm}
\includegraphics[width=8.5cm,angle=0.]{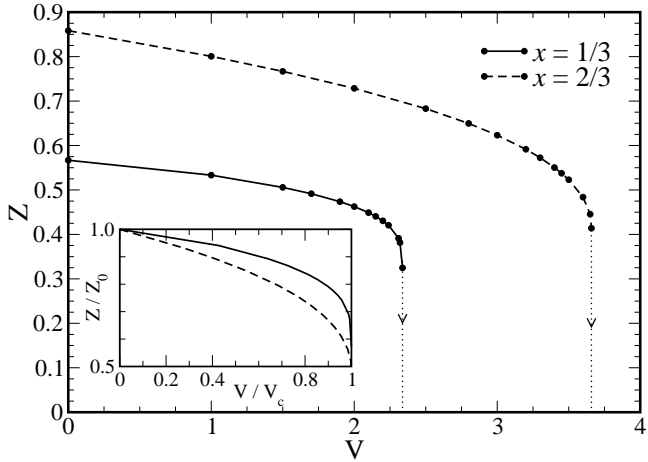}
\end{center}
\caption{
The quasiparticle weight $Z$ for $J=0.2$ evaluated on the Fermi surface as a 
function of $V$ for $x = 1/3$ (solid line)
and $x = 2/3$ (dashed line).
With increasing $V$ both curves show that, first, $Z$ decreases slowly and, it 
drops to zero close to $V_c$ ($Z \rightarrow 0$ for $V \rightarrow V_c$).
Dotted lines are guides for the eyes which show the mentioned tendency (see 
text for discussions).
For $V=0$ the quasiparticle weight for $x=1/3$ is smaller than for $x=2/3$ 
which means that correlations are stronger for small than for large $x$.
Inset: The QP weight divided by the QP weight at $V=0$ ($Z/Z_0$) versus 
$V/V_c$ for $x=1/3$ (solid line) and $x=2/3$ (dashed line). 
With increasing $V$, the decreasing of $Z$ is somewhat faster for $x=2/3$ than for 
$x=1/3$.
See text for discussions.
}
\label{fig:ZvsV}
\end{figure}

The self-energy allows the calculation of the quasiparticle (QP) weight
$Z=1/(1-\frac{\partial \Sigma}{\partial \omega})$ at the Fermi level. In 
Fig.\ \ref{fig:ZvsV}, the QP weight is plotted as a function of $V$ for 
$x=1/3$ (solid line) and for $x=2/3$ (dashed line).
The Fermi vectors are located in the $\Gamma-K$ direction of the 
Brillouin zone for both,  
$x=1/3$ and $x=2/3$. 
As we will 
discuss below, our self-energy is very isotropic on the Fermi surface (FS) 
such that Fig.\ \ref{fig:ZvsV} is representative for all points on the FS.
In both cases, when $V$ approaches 
the corresponding $V_c$,
the QP weight decreases suggesting that 
it tends to zero at $V_c$ (see dotted arrows).
Below, the analysis of spectral weight transfer at small energy will give 
additional support for this result.

The behavior presented in Fig.\ {\ref{fig:ZvsV}} indicates the breakdown of 
the Fermi liquid (FL) at $V_c$.
Therefore, our calculation suggests a metal insulator transition 
where a rather 
isotropic gap
should exists for $V>V_c$. Our observation is 
based
on the fact 
that for $V<V_c$ the QP weight $Z$ is very isotropic on the FS 
such that
the FL 
breaks 
down
for all points on the FS.

In both curves, when $V$ increases the QP weight decreases, first 
relatively slowly,
and then drops to zero close to $V_c$.
Therefore, near the CDW instability the QP carries very small weight 
making the electronic dynamics very incoherent.
To clarify this point better we will present
below
results for the spectral functions. 

The QP weight for $V=0$ (far from the CDW instability) is already 
reduced from one. This reduction is not related 
to
the CDW instability but it is due to electronic correlation effects 
which our method is able to capture in the pure $t$-$J$ model.
In fact, Fig.\ \ref{fig:ZvsV} shows that
the QP weight for $x=1/3$ is smaller than for $x=2/3$, as one expects, since 
correlation effects should be weaker for a dilute system.
For $x\rightarrow 1$ our method predicts $Z \rightarrow 1$ which corresponds 
to the uncorrelated limit (empty band). On the other hand, in both cases, 
$Z$ strongly decreases on approaching the charge order intstability.

\subsection{Self-energy}

The discussion above indicates that the HFL is strongly modified near the CDW 
instability,
where the carriers are renormalized by interacting with
soft charge fluctuations. In order to discuss this point in 
detail, we will present
results for the self-energy $\Sigma({\bf k}, \omega)$.

\begin{figure}
\begin{center}
\setlength{\unitlength}{1cm}
\includegraphics[width=8.5cm,angle=0.]{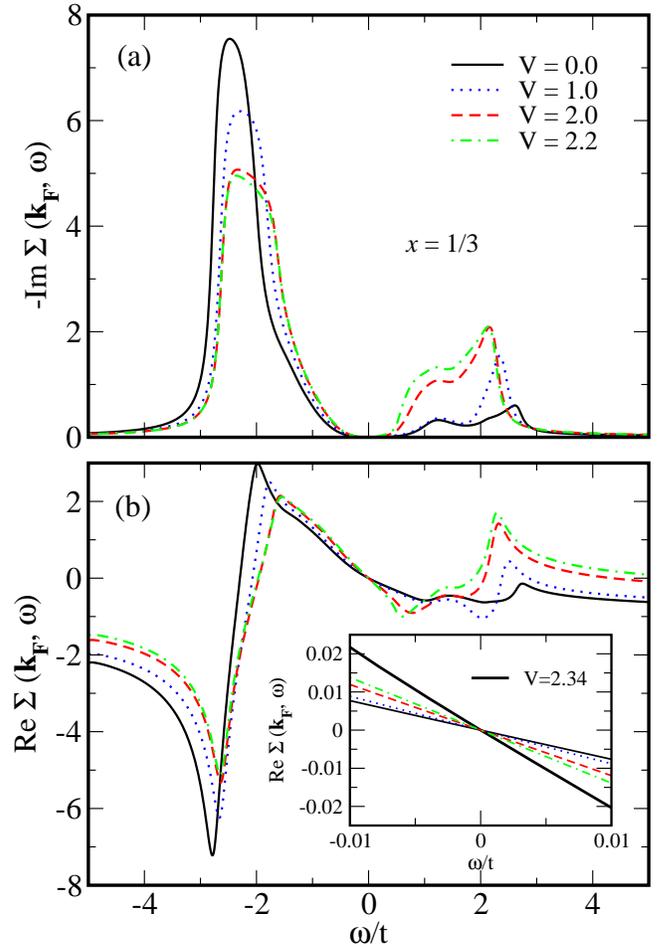}
\end{center}
\caption{(color online)
Self-energy $\Sigma ({\bf k_F}, \omega)$ for $x = 1/3$ and $J=0.2$ for 
several 
values of
$V$ approaching $V_c 
\sim 2.35
$.
(a) $-Im \, \Sigma ({\bf k_F}, \omega)$ versus $\omega$. It presents large 
structures at
energies of the order of $t$ and shows a $\sim \omega^2$ behavior around 
$\omega = 0$.
With increasing $V$ a redistribution of the weight takes place.
(b) $Re \, \Sigma ({\bf k_F}, \omega)$ versus $\omega$ for the same 
values of
$V$ 
as
in panel (a).
Inset shows that, with increasing $V$, the slope of 
$Re \, \Sigma$ near $\omega = 0$ increases
leading to 
a decrease
of the quasiparticle weight.
}
\label{fig:sigma}
\end{figure}

In Fig.\ \ref{fig:sigma}, self-energy results at ${\bf k}={\bf k_F}$ are 
presented for $x=1/3$ and several 
values of $V$ approaching $V_c$. Panel (a) shows 
$-Im \, \Sigma$ as a function of frequency. As discussed in Ref.\ 
[\onlinecite{bejas06}], for $V=0$,  $Im \, \Sigma$ is strongly asymmetric 
with respect to $\omega=0$ and behaves as $\sim \omega^2$ at small 
$\omega$'s. The self-energy presents large contributions 
at large energy of 
the order of $t$
below the Fermi energy.
With increasing $V$ a redistribution of the weight takes place and 
$Im \, \Sigma$ develops structures at 
energies close to but above the Fermi energy.
As $Im \, \Sigma$ and $Re \, \Sigma$ are related 
to
each other by a Kramers Kronig 
transformation, the changes in $Im \, \Sigma$ are reflected in 
$Re \, \Sigma$ (panel (b)).
For instance the slope of $Re \, \Sigma$ at $\omega=0$ increases with 
increasing $V$ leading to a decrease
of the QP weight discussed in Fig.\ \ref{fig:ZvsV}.
The inset in panel (b) shows this behavior in 
a smaller scale near $\omega = 0$.
While for $V \leq 2.2$ a gradual change in the slope is observed, very close to
$V_c$, a stronger increase is observed, as shown by the thick solid line
corresponding to $V=2.34$. 

\subsection{Spectral functions and total density of states}
\begin{figure}
\begin{center}
\setlength{\unitlength}{1cm}
\includegraphics[width=8.5cm,angle=0.]{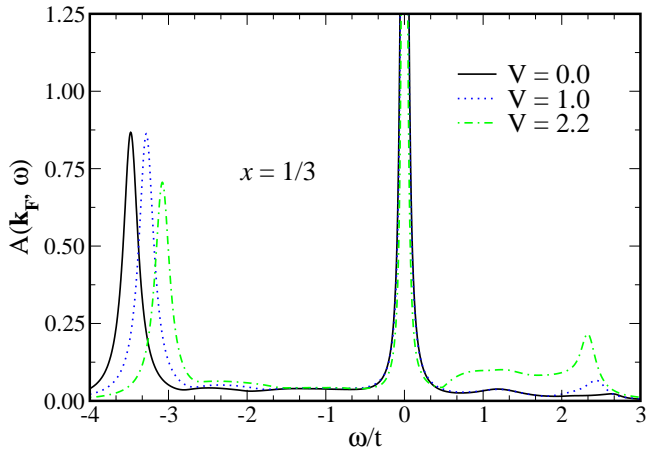}
\end{center}
\caption{(color online)
Spectral function $A({\bf k}_F, \omega)$ versus $\omega$ for $J=0.2$ and 
several values of $V$
approaching $V_c
\sim 2.35
$.
For $V=0$ (solid line), pure $t$-$J$ model correlation effects leads to a 
quasiparticle peak with reduced weight $Z \sim 0.56$ and incoherent 
structures at positive and negative energies.
With increasing $V$, while the QP weight decreases (Fig.\ \ref{fig:ZvsV}), 
weight is transfered to the range $\omega \sim 0.5t - 2.5t$.
}
\label{fig:spec}
\end{figure}

Using $Im \, \Sigma$ and $Re \, \Sigma$, the spectral function 
$A({\bf k},\omega)$ can be calculated as usual.
Figure \ref{fig:spec} presents results for $A({\bf k}_F, \omega)$ for several  
values of $V$. 
The peak at $\omega=0$ corresponds to the QP while the other features, for 
instance at $\omega \sim -3t$ and $\omega \sim 2.5t$, are of incoherent 
character. For $V=0$ (solid line) the QP weight is $Z\sim 0.56$ 
(Fig.\ \ref{fig:ZvsV}) and the incoherent structure is mainly concentrated in 
the pronounced feature at $\omega \sim -3.5t$.
Increasing $V$ up
to $V=2.2$ (dotted dashed line)  
the QP weight decreases ($Z \sim 0.42$) 
and, at the 
same time, spectral weight appears
at $\omega \sim 0.5t - 2.5t$ (Fig.\ \ref{fig:spec}). 
In addition, with 
increasing $V$, the incoherent structure at $\omega \sim -3.5t$ moves to the 
right while loses some weight.

In Fig.\ \ref{fig:spec} it was assumed, supported by the Luttinger theorem, 
that ${\bf k_F}$ does not change with the interaction $V$.
On the other hand, if we wish to compare with $ARPES$ experiments (see 
Sec.\ \ref{sec:cobaltates}), our input is a tight
binding dispersion which reproduces the measured FS which already contains 
the interactions.
The comparison between our approach and Lanczos diagonalization presented in 
Ref.\ [\onlinecite{bejas06}] gives an additional support fort this assumption 
(see also Ref.\ [\onlinecite{merino03}]).

\subsection{Physical origin of the self-energy renormalizations: $\alpha^2F(\omega)$}
In the following we discuss the interaction of the bosonic excitations 
described by $D_{RR}$ (see eq.\ (\ref{chiv1})) with electrons that lead to 
the self-energy renormalizations discussed in the previous section.
In many body theory
\cite{mahan} the quantity which contains this information is 
$\alpha^2 F(\omega)$ where, $F(\omega)$ is the
density of states of a boson which interacts with the fermions with strength 
$\alpha$.
At $T=0$ \cite{norman06},
\begin{equation} \label{eq:a2F}
 Im \, \Sigma(\omega) = \int_0^\omega d\Omega \; \alpha^2F(\Omega) \; ,
\end{equation}
\noindent i.e. $\alpha^2F(\omega)=\partial Im \, \Sigma(\omega) / \partial \omega$.
In eq.\ (\ref{eq:a2F}) the average over all momentum transfer 
${\bf q} = {\bf k}_1 - {\bf k}_2$, where ${\bf k}_1$ and ${\bf k}_2$ are two 
Fermi vectors, is assumed \cite{mahan}.

\begin{figure}
\begin{center}
\setlength{\unitlength}{1cm}
\includegraphics[width=8.5cm,angle=0.]{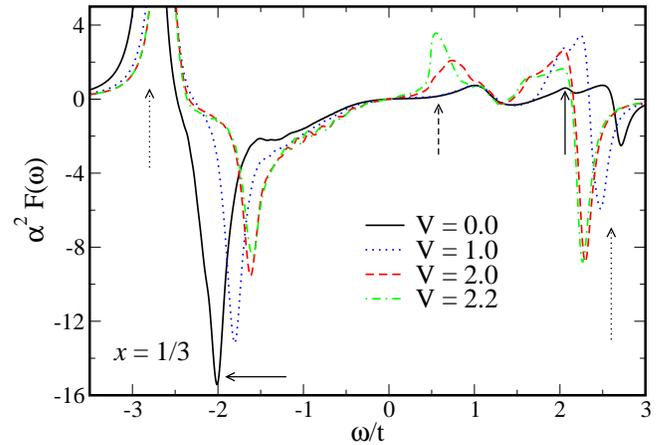}
\end{center}
\caption{(color online)
$\alpha^2F(\omega)$ for the same values of $V$ as in
Fig.\ \ref{fig:sigma}.
The features marked with solid arrows are related with collective charge 
fluctuations which become soft
with increasing $V$ and, at the same time, $\alpha^2F(\omega)$ acquires 
structures at low energies (dashed arrow)
which lead to the decreasing of $Z$ (see text).
}
\label{fig:a2F_deriv}
\end{figure}

In Fig.\ \ref{fig:a2F_deriv} we have plotted $\alpha^2F$ for the same 
values of $V$ as in Fig.\ \ref{fig:sigma}.
As we mentioned above, our self-energy is very isotropic on the FS such that
$\alpha^2F$ evaluated at ${\bf k_F}$ is representative for the average over 
the FS.

\begin{figure}
\begin{center}
\setlength{\unitlength}{1cm}
\includegraphics[width=8.5cm,angle=0.]{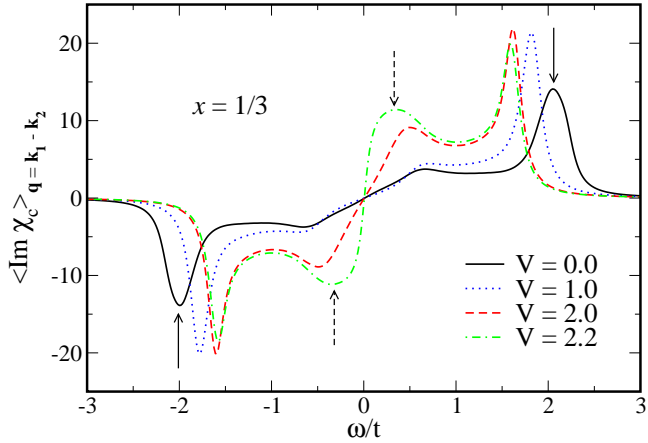}
\end{center}
\caption{(color online)
Average of $Im \, \chi_c ({\bf q}, \omega)$ for all ${\bf q}$ such that 
${\bf q} = {\bf k_1} - {\bf k_2}$
with ${\bf k}_1$ and ${\bf k}_1$ on the FS. With increasing $V$, the 
collective charge peaks (solid arrows)
become soft and at the same time a soft dynamics (dashed arrows) appears at 
low energies. The close correlation
between these features and those discussed in Fig.\ \ref{fig:a2F_deriv} 
shows that the charge
dynamics is the main cause for the reduction of $Z$ near charge 
order.
}
\label{fig:a2F}
\end{figure}

Let us discuss first the case for $V=0$ (solid line). We can see two 
structures, at $\omega \sim -2t$ 
and $\omega \sim 2t$ (solid arrows) which are related with collective charge 
excitations (see below).
The features at $\omega \sim -3t$ and $\omega \sim 2.5t$ (dotted arrows) are 
due to the fact that $Im \, \Sigma$ (Fig. \ref{fig:sigma}) is concentrated in 
a finite range of energy (between the borders $\sim -4t$ and $\sim 3t$) 
so that, when performing the derivative 
$\partial Im \, \Sigma(\omega) / \partial \omega$ to obtain 
$\alpha^2 F(\omega)$, these two features are created at these two borders.
In Fig.\ \ref{fig:a2F} we have plotted the average of the 
$Im \, \chi^c ({\bf q}, \omega)$ for all ${\bf q}$ such 
that 
${\bf q}={\bf k_1}-{\bf k_2}$ where ${\bf k_1}$ and ${\bf k_2}$ are two Fermi 
vectors.
For $V=0$ (solid line), the well defined peaks (solid arrows) at 
$\omega \sim -2t$ and $2t$ are due to collective charge excitations while 
tails are due to the particle-hole continuum (see solid line in 
Fig.\ \ref{fig:sus33}).
Clearly, the two structures marked with solid arrows in 
Fig.\ \ref{fig:a2F_deriv} are correlated with the corresponding ones in 
Fig.\ \ref{fig:a2F} showing that these structures in $\alpha^2F$ are related 
to collective charge fluctuations while tails, are due to the particle-hole 
continuum.
The reason why $\alpha^2F(\omega)$ does not trace exactly the average of the 
charge correlations of Fig.\ \ref{fig:a2F} is the following.
$\alpha^2F$ contains information of both, the density of states of the 
interacting boson and its coupling with fermions but, in a mixed form.
For instance, the only way that $\alpha^2F$ follows exactly the same shape of 
the boson density of states occurs if the coupling does not depend on 
$\omega$ or momentum.
This condition is not satisfied in our case which can be seen by looking at 
the expression (\ref{eq:SigmaIm0}) for $Im \, \Sigma$.
On the other hand, only in the first term of eq.\ (\ref{eq:SigmaIm0}) the 
charge-charge correlation is explicitly present.
The other terms which contain $D_{\lambda R}$ and $D_{\lambda \lambda}$ are 
proper of our method, and they are due to the
nondouble occupancy constraint.

With increasing $V$, the features marked with
solid arrows in Fig.\ \ref{fig:a2F_deriv} become soft and, at the same time, 
spectral weight appears at lower
energies (dashed arrows). This behavior is clearly 
correlated with that depicted in Fig.\ \ref{fig:a2F} which shows the soft 
charge dynamics discussed
in Sec.\ \ref{sec:CDW} (see Fig.\ \ref{fig:sus33}).

It is well known \cite{mahan} that
\begin{equation}\label{eq:al}
-\frac{\partial Re \, \Sigma}{\partial \omega} = 
\int d\omega \frac{\alpha^2F(\omega)}{\omega} \; .
\end{equation}
Therefore, the presence of low energy excitations in $\alpha^2F$ 
leads to an increase of $\partial Re \, \Sigma/\partial\omega$ and 
hence, to a decrease of the QP weight $Z$.
Notice that near $V_c$, $\alpha^2 F(\omega)$ shows, besides 
the low energy 
features, structure at high energy in agreement with the behavior presented 
by charge correlations (Fig.\ \ref{fig:sus33} and Fig.\ \ref{fig:a2F}).
The fact that not all the spectral weight is concentrated at low energy is 
the cause for the slow decrease of $Z$. 
The inset of Fig.\ \ref{fig:ZvsV} shows that $Z$ decreases 
somewhat faster for $x=2/3$ 
than 
for $x=1/3$ which is consistent with the fact that softening of the charge 
collective mode is more clear for $x=2/3$ (see Figs.\ \ref{fig:sus66} and 
\ref{fig:sus33}).

To conclude, we think that we have given clear
arguments which show that the charge collective soft modes near the CDW are 
responsible for the incoherent motion of the carriers near 
charge 
ordering.
In addition, the charge soft modes reach $\omega = 0$ at $V_c$ which means that
$-\partial Re \, \Sigma / \partial \omega \rightarrow \infty$ 
(eq.\ \ref{eq:al}) and
then $Z \rightarrow 0$ as indicated in Fig.\ \ref{fig:ZvsV}.

\subsection{E-k structures}
\begin{figure}
 \begin{center}
  \setlength{\unitlength}{1cm}
  \includegraphics[width=8.5cm,angle=0.]{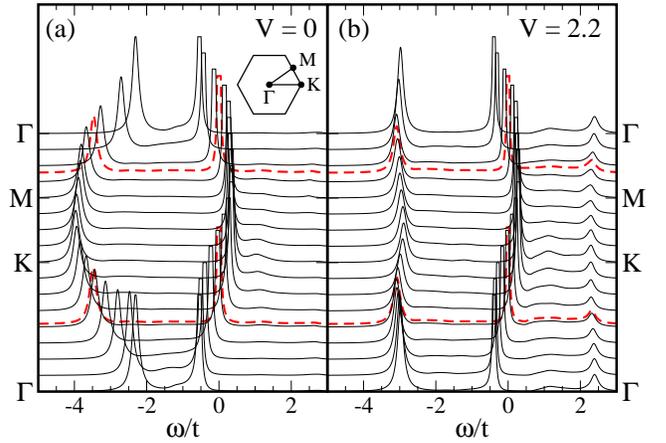}
 \end{center}
 \caption{(color online)
Spectral functions at $x = 1/3$ in the main directions of the Brillouin zone 
for (a) $V=0$ and (b) $V=2.2$.
The peak near $\omega = 0$ is the coherent quasiparticle peak and the other 
structures are of incoherent
character.
For $V=2.2$ the coherent QP band becomes flatter and an incoherent band 
appears in the energy
range $\omega \sim 0.5t - 2.5t$.
The QP peak has been cut for clarity purposes.
}
\label{fig:disper}
\end{figure}

When discussing the spectral functions we have shown the presence of coherent 
QP peaks and incoherent structures.
In Fig.\ \ref{fig:disper} we show the energy position of the main visible 
structures for $V=0$ and $V=2.2$ at $x=1/3$ along
the main directions of the Brillouin zone (inset in Fig.\ \ref{fig:disper}).
Dashed lines mark the spectral functions at ${\bf k_F}$.
For $V=0$ (panel (a)), we can see that the QP band (near the FS) and a 
dispersing incoherent band at $\omega \sim -3t$ dominate the spectra.

When $V$ increases to $V=2.2$ (panel (b)) the QP coherent band becomes less 
dispersing consistently with the reduction of the QP weight.
The incoherent band at negative $\omega$ becomes also less dispersing. 
Interestingly, an incoherent band is formed in the region $\sim 0.5t-2.5t$.
Notice that no kink is showed by the coherent QP band.
This is because the kink, if it exist, should be located at energies of the 
order of the interacting boson (which in this case is related to the soft 
mode) and, at $V=2.2$, it is still located at energies of $\sim 0.2t$ 
(Fig.\ \ref{fig:sus33}) which is of the order of the QP bottom energy.

\section{Comparison with cobaltates} \label{sec:cobaltates}
In previous sections we have discused one-particle spectral predictions of 
the large-$N$ approach for the $t$-$J$-$V$ model on the triangular lattice 
without making emphasis on any particular system.
In this section we will discuss previous results and their 
possible contact 
with cobaltates.
First principle calculations in cobaltates \cite{singh03} predict, besides a 
large hole Fermi surface around $\Gamma$, the existence of six small pockets 
near the corners of the Brillouin zone. Until now, all ARPES experiments both, 
in hydrated\cite{qian06_97,shimojima06} and unhydrated cobaltates 
\cite{qian06,hasan04,yang04,dqian06p,mhasan06p,kuprin06,yang05} do not show 
the presence of these small pockets.
The absence of pockets in ARPES can be understood by the renormalization of 
the bands due to strong electronic correlations \cite{zhou05}. 
In fact, a reduction of the bandwidth 
and large electronic effective masses are observed in ARPES
\cite{dqian06p,mhasan06p}.
In addition, in a very recent work \cite{vaulx07} it was 
pointed out
that cobaltates are close to a Mott insulator in the 
limit $x \rightarrow 0$

Following these ideas, a one-band $t$-$J$ model on the triangular lattice, as 
proposed in 
the 
present paper, seems to be appropriated for studying cobaltates.
For this purpose,
$\tilde{c}^\dag_{i\sigma}$ and $\tilde{c}_{i\sigma}$ in eq.\ (\ref{eq:Hc}) 
must be associated with the 
creation and destruction of holes respectively, as in 
Refs.\ [\onlinecite{motrunich04,foussats05,motrunich03}].
Then, in this context, the doping $x$ is the electron doping away from 
half-filling.
Considering the bare hoping $t \sim 150meV$\cite{singh00} and 
$U_{dd} \sim 3.0-5.5 eV$ \cite{chainani04} we obtain $J/t \sim 0.1-0.2$, 
similarly to the value used in previous sections.
In addition, we will mainly focus the comparison with our 
results for 
$x=1/3$, which is close to the doping where the maximum 
superconducting 
critical temperature 
$T_c$ is found in cobaltates.
For $x<0.5$ our results predict a Pauli paramagnet as observed in cobaltates. 
On the other hand, as cobaltates exhibit a Curie-Weiss behavior in the region 
$x>0.5$, we will not relate our results in $x=2/3$ with experiments.
However, it is important to mention that it is not clear if this Curie-Weiss 
behavior is due to the fact that electronic correlations are more important 
for $x>0.5$ or other effects, as for instance $Na$ order,  
are predominant (see Ref.[\onlinecite{schulze07}] and 
references their in).

According to results in previous sections we have two factors which reduce 
the Fermi velocities from the bare one.
The first one occurs already at mean field level where the effective hopping 
is $(x/2)t$ instead of $t$.
The second one, which is obtained only after evaluating fluctuations 
beyond mean field is the quasiparticle weight $Z$. Therefore, the 
Fermi velocity is $v_F \sim (x/2)\; Z \; v_F^{LDA}$ where $v_F^{LDA}$ is the 
bare Fermi velocity extracted from LDA calculations.
Using the parameters for cobaltates and our results for $x=1/3$ up to 
$
0.9 
V_c$ we obtain $v_F \sim 0.21eV\AA$.
This value is close to the value $v_F = (0.30\pm0.05) eV\AA$ found in 
Refs.\ [\onlinecite{mhasan06p,dqian06p}].
This small value was considered\cite{mhasan06p} as an indication for the 
presence of strong correlations and, the ratio between $T_c$ and the Fermi 
velocity was found to be close to the same ratio for cuprates.
In agreement with these experiments, our calculation shows also a rather 
isotropic Fermi velocity over the BZ (see Fig.\ \ref{fig:disper}).
Another interesting and qualitative agreement between experiments and our 
finding is the following:
in Ref.\ [\onlinecite{mhasan06p}] it was found that the scattering rate 
behaves as $\sim \omega^2$ near the Fermi surface which agrees with results 
in panel (a) of Fig.\ \ref{fig:sigma}.

In our previous papers\cite{foussats05,foussats06}, for describing 
superconductivity in cobaltates we proposed, like other works 
\cite{motrunich04,motrunich03,tanaka04}, that the system is close to 
$\sqrt{3} \times \sqrt{3}$ charge order.
The charge order scenario is supported by some 
experiments\cite{shimojima05,lemmens05,qian06}.
In addition, ARPES experiments \cite{dqian06p} show a Fermi surface topology, 
very close to our FS, which favors charge instabilities.
We have proposed that the interplay between electronic correlations and 
phonons are relevant for describing
superconductivity with triplet NNN-$f$ pairing as predicted by some experiments
\cite{fujimoto04,kato06,ihara05,kanigel04,higemoto04}.
For this scenario we need that, with increasing water content, the system 
comes closer to charge ordering.
Recent ARPES experiments\cite{qian06_97,shimojima06} have shown that the 
Fermi velocity do not vary very much with the water content.
These experiments seem to be in contradiction with our result of Fig.\ 
\ref{fig:disper} where it is possible to see that very close to the charge 
instability the band dispersion becomes flat.
However, we think that it is possible that for the water content reached in 
the experiments the system is still far from the static charge order state 
i.e., in the regime where $Z$ does not change very much.
This could explain the reason why NMR\cite{mukhamedshin05} does not show the 
charge order because it is fluctuating faster than the time-scale accessible 
by NMR.
We remember that for obtaining a robust description of superconductivity we 
do not need to place the system very close to the charge order being enough a 
$V\sim 80-90\%V_c$ \cite{foussats05,foussats06}. However, for $V > 0.9 V_c$, $Z$ is strongly reduced and 
two competing 
effects can be expected. From one side, the decreasing of 
$Z$ may cause an 
increasing of the electronic density of states and then, 
a reinforcement 
of superconductivity. On the other side, when $Z$ 
decreases the quasiparticle 
losses coherence thus, superconductivity may be diminished. 
A careful study 
of this competition requires the calculation of 
superconducting pairing in 
${\cal O}(1/N^2)$.

It is interesting to compare our results with a very recent ARPES experiment 
on a series of $Bi$ cobaltates \cite{brouet}.
These compounds contain the same triangular $Co$ planes as $Na$ cobaltates.
Besides a flat QP band with similar Fermi velocity as in 
$Na$ cobaltates, incoherent features were observed at high energy in the $Bi$ 
cobaltates.
In that paper it was discused that these
incoherent features are created with weight transfered from the QP band, a 
mechanism similar to that discussed in Sec.\ \ref{sec:OPEP} (notice the 
analogy, for instance, between Fig.\ 2 of Ref.\ [\onlinecite{brouet}] and our 
Fig.\ \ref{fig:disper}).
On the other hand, the center of the incoherent band in Fig.\ 2 of Ref.\ 
[\onlinecite{brouet}] is at $\sim -0.5$ eV, close to the center of our 
incoherent band (Fig.\ \ref{fig:disper}) which, using $t = 150$ meV is about 
$-0.45$ eV. Similar high energy features were recently 
observed in cuprates (see Ref.\ [\onlinecite{brouet}] and references their in) and, 
the 
present approach was used 
\cite{greco07} for discussing those features occuring in 
the
overdoped side where, as mentioned before, our method is expected to be reliable.

Optical conductivity experiments \cite{hwang05,wu06}  
show a different behavior to usual 
metals where, besides a Drude peak at low energies, a broad absorption 
centered around $\sim 250$ meV
is observed and interpreted as a pseudogap\cite{wu06} behavior. Similar 
features are observed in organic materials and discussed in term of the 
charge order proximity \cite{dressel03}.
From our results in Fig.\ \ref{fig:spec} we expect that, near the charge 
order, transitions from the Drude peak to the broad structure formed between 
$\sim 75-375$ meV (where we used $t=150$ meV) are possible leading to the 
optical absorption as in the experiments.

Recently, optical conductivity experiments in organic materials 
\cite{drichko06} have also shown an anomalous increasing of the effective 
mass with temperature.
In Ref.\ [\onlinecite{merino06}], a reentrant behavior in the $T$-$V$ phase 
diagram, as the one discussed in our Fig.\ \ref{fig:TV}, was found to be 
responsible for the increase in effective mass. 
If for cobaltates $0.7V_c < V < V_c$, we think that an increase in 
effective mass with temperature
can also be expected in these materials.
From Fig.\ \ref{fig:TV}, depending on the value of $V$, the effective mass 
increase may occurs for $T < 800$ K.
It will be interesting to perform this kind of experiments.

\section{Conclusion and discussions}
In this paper, the $t$-$J$-$V$ model at finite density  was studied on the 
triangular lattice.
$V$ represents the Coulomb interaction between nearest neighbors and it is 
the parameter responsible for triggering a charge order phase for $V>V_c$. 
$V_c$ is the critical 
Coulomb repulsion whose numerical value depends on doping 
(Fig.\ \ref{fig:fase}).
Studying charge correlation functions, it was shown that near $V_c$ charge 
dynamics becomes soft and, at $V=V_c$, charge modes collapse to zero 
frequency freezing a $\sqrt{3} \times \sqrt{3}$ CDW phase 
(Figs.\ \ref{fig:sus66} and \ref{fig:sus33}).

By evaluating fluctuations beyond 
the
mean field level, one-particle spectral properties were computed.
Far away from charge ordering,
i.e.\ for zero or small $V$, the QP weight $Z$ is reduced from one due to 
pure $t$-$J$ model correlations effects.
With increasing $V$, $Z$ decreases slowly 
until $V \sim 90\%$ of $V_c$. \
Beyond this value $Z$ decreases faster approaching zero at $V_c$ 
(Fig.\ \ref{fig:ZvsV}).
Therefore, near $V_c$ the electronic 
dynamics 
becomes very incoherent.
For $V<V_c$, the scattering rate behaves as $Im \, \Sigma \sim \omega^2$ 
(Fig.\ \ref{fig:sigma}(a)) which is characteristic for a Fermi liquid 
behavior. In addition, when the system approaches charge ordering,
$Im \, \Sigma$ accumulates spectral weight at low energy.
Due to this fact, the slope of $Re \, \Sigma$ at $\omega=0$ increases 
(Fig.\ \ref{fig:sigma}(b)) leading to the decrease of $Z$ discussed above.

From $Im \, \Sigma$ and $Re \, \Sigma$ the spectral function $A(k,\omega)$  
was calculated. 
With increasing $V$, approaching $V_c$, the most important changes are 
present at 
energies close but larger than the Fermi energy ($\omega=0$).
While the weight of the QP peak decreases ($Z$ decreases), spectral weight 
appears at small positive energy in the range $\sim 0.5t-2.5t$ 
(Fig.\ \ref{fig:spec}).

The calculation of $\alpha^2 F(\omega)$ 
allows to identify the excitations responsible for the self-energy effects.
Computing $\alpha^2F(\omega)$ as a function of $V$ approaching $V_c$, it was 
shown that its behavior can be, without ambiguity, correlated with charge 
spectra (Figs.\ \ref{fig:a2F_deriv} and  
\ref{fig:a2F}).
Therefore, the charge soft modes, responsible for the charge instability, 
lead to an increase of 
$\alpha^2 F$ at small frequencies with the corresponding 
decrease of $Z$.

In Sec.\ \ref{sec:cobaltates}, we mainly confronted our results with recent 
$ARPES$ experiments in cobaltates. We found our results to be in  
agreement with experiments presented in Refs.\ 
[\onlinecite{dqian06p,mhasan06p,brouet}].

Finally, we would like to point out that
while our results were obtained for a purely 2D system, we expect
only quantitative differences in our results when passing to a 
corresponding anisotropic 3D system, since we are dealing here
with a spontaneous symmetry breaking of a discrete symmetry, as corresponds 
to a commensurate CDW. Certainly critical exponents may change, but they are 
beyond the scope of our treatment.

{\bf Acknowledgements} A. G. thanks to M. Dressel, N. Drichko and J. Merino for interesting discussions.


\begin{thebibliography}{10}
\bibitem{aclara1}
$\gamma_q$ was corrected (a factor $2$ was removed) from previous versions.
\bibitem{aclara2}
A typing error in the second term of the first component of $h_a$ in Ref.[\onlinecite{bejas06}]
was corrected now in eq.(\ref{eq:ha}).
\end{thebibliography}
\end{document}